\documentclass[referee]{raa}            

\usepackage{graphicx,times}             
\usepackage{natbib}
\usepackage{amssymb,amsmath}
\bibpunct{(}{)}{;}{a}{}{,}

\usepackage[a4paper=true,dvipdfm=true,pagebackref=true]{hyperref}
\hypersetup{colorlinks = true, linkcolor = green, anchorcolor = red,
 citecolor = blue, filecolor = red, pagecolor = red, urlcolor = red}

\begin{document}

   \title{Search for the brightest stars in galaxies outside the Local Group
\,$^*$
\footnotetext{$*$Based on observations with the NASA/ESA Hubble Space Telescope, obtained at the
Space Telescope Science Institute, which is operated by AURA, Inc. under contract No. NAS5--26555.
These observations are associated with proposal 10402,10354,11578,14716,15654.}
}

   \volnopage{Vol.0 (20xx) No.0, 000--000}     
   \setcounter{page}{1}          
\author{N.A.~Tikhonov\inst{1}, O.A.~Galazutdinova\inst{1}, O.N.~Sholukhova
\inst{1}, A.~Valcheva\inst{2}, P.L.~Nedialkov\inst{2}, O.A.~Merkulova\inst{3}   }

   \institute{Special Astrophysical Observatory, Nizhnij Arkhyz,
 Karachai-Cherkessian Republic,
    Russia 369167; {\it ntik@sao.ru}\\
                   \and    
	Department of Astronomy, Sofia University, Sofia, Bulgaria, 1504
 Sofia, 15 Tsar Osvoboditel Blvd. \\
 \and
 Astronomical Institute, St.Petersburg State University,
 Universitetskii pr. 28, St.Petersburg, 198504 Russia
\vs\no
   {\small Received~~2019 November 15; accepted~~20xx~~month day}}
{\abstract{This paper shows a technique for searching for bright massive stars in
galaxies beyond the Local Group. To search for massive stars, we used
the results of stellar photometry of the Hubble Space Telescope images
using the DAOPHOT and DOLPHOT packages. The results of such searches are
shown on the example of the galaxies DDO\,68, M\,94 and NGC\,1672. In the galaxy
DDO\,68 the LBV star changes its brightness, and in M\,94 massive stars can be
identified by the excess in the H${\alpha}$ band. For the galaxy NGC\,1672,
we measured the distance for the first time by the TRGB method, which made
it possible to determine the luminosities of the brightest stars, likely
hypergiants, in the young star formation region. So far we have performed stellar photometry of HST images of 320~northern sky galaxies located at a
distance below 12~Mpc. This allowed us to identify 53 galaxies with probable hypergiants. Further photometric and spectral observations of these galaxies are planned to search for massive stars.}}

   \authorrunning{N.A.~Tikhonov et al.}           
   \titlerunning{Search for the brightest stars in galaxies outside the
 Local Group } 

   \maketitle

\section{Introduction}          
\label{sect:intro}

One of the main tasks of modern astrophysics is to determine the upper limit
 of the mass of a star. The interest lies in the fact that massive stars
 evolve quickly and the final stage of their evolution can be a supernova
 burst, formation of a black hole or a neutron star, or even a merger of
 such relativistic objects with the release of huge amount of energy. The
 evolutionary path of a star depends on its initial mass. Models of stellar
 evolution suggest that the initial masses of stars can reach up to 500~solar
 masses \citep{YHN2013,SM2017} and more, but so far, only the
 stars of much smaller mass have been discovered  \citep{CSH2017}.  The
 question of the fidelity of theoretical models can only be solved by
 searching for and studying massive stars in galaxies of various masses.

In our Galaxy, massive stars are observed in clusters located in the galactic
 plane, where the extinction of light can be very high \citep{EML2004}.
 To increase the breadth of the search and solve the problem with the
 uncertainty of light extinction, one should go beyond the Galaxy and search
 for massive stars in other, not too distant galaxies \citep{HDH2017,SFV2018}.
  Using the Hubble Space Telescope, numerous galaxies
 outside the Local Group are accessible for the study (up to distances of
 15--20\,Mpc).

\section{Search for candidates for massive stars}
\label{sect:candidates}

Spectral observations make it possible to determine the type of star and
 evaluate its physical parameters, but such observations are not suitable
 for search tasks because of the large number of candidates for massive stars
 and high time expenditures to obtain the results. Therefore, to identify
 possible candidates for massive stars (hypergiants), photometric methods
 are used, as they are faster and more accessible on all telescopes.

It seems that finding massive stars, which are also the brightest stars in
 galaxies, is not difficult, but this opinion is incorrect.
 In many low-mass galaxies, there is simply no such stars.
 The lifetime of these stars is very small (several million years), and the
 frequency of their appearance in galaxies is small. According to the law of
 mass distribution of Salpeter, thousands of stars of smaller masses should
 be born per star with a mass of 100 Solar masses.
Even in our giant Galaxy, with numerous of star formation regions, only a
 few stars with masses of 150---300 Solar masses
are known: {HD15558A \citep{GM1981}, Eta Carinae \citep{G2015}}.

\begin{figure}[h]
   \centering
   \includegraphics[width=10cm, angle=0]{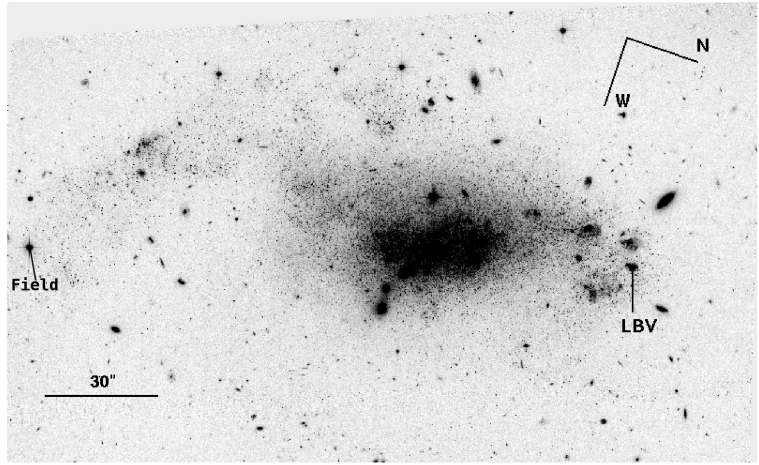}
   \caption{Dwarf metal-poor galaxy DDO\,68 in the F606W ($V$) filter with the LBV star, marked by a
 circle and a bright blue field star belonging to our Galaxy.}
   \label{Fig1}
   \end{figure}

\begin{figure}[h]
   \centering
   \includegraphics[width=5cm, angle=0]{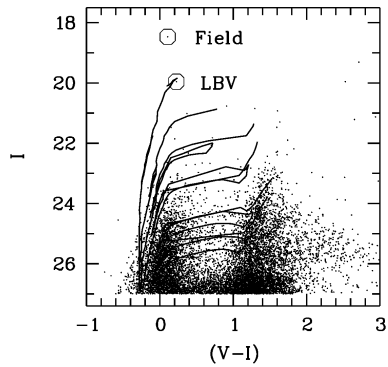}
   \caption{CM-diagram of DDO68 stars and isochrones of age from 3~Myr to
100 Myr and metalicity $Z = 0.001$ from \citet{BBC1994}. The position
 of the LBV star and the foreground star is marked. }
   \label{Fig2}
   \end{figure}

The second factor that complicates the search is the superposition of stars
 of our Galaxy on the studied fields. An example is given by the metal-poor
 dwarf galaxy DDO\,68 (Figure~\ref{Fig1}), the distance to which is 12$\pm$0.3\,Mpc
 \citep{TGL2014}. In this galaxy, a bright massive LBV (Luminous Blue Variable)
 star is known \citep{PMP2017}, which is marked on the image of the galaxy (Figure~\ref{Fig1}).
 { On the Hertzsprung-Russell diagram (CM-diagram) DDO\,68
 (Figure~\ref{Fig2}), this star is very bright (according to our measurements,
 $M_V = -10\,.\!\!^{\rm m}26$). However, this CM-diagram contains another brighter blue
 star, also visible on the body of the galaxy (Figure~\ref{Fig1}). As shown by
 spectral observations with 6-m telescope BTA, this bright blue star has no
 emission lines, and its radial velocity is low compared to the DDO\,68. This
 indicates that it is not a massive star DDO\,68, but belongs to our Galaxy.
 Isochrones of \citet{BBC1994} overploted on the DDO\,68 CM-diagram also show
 that the bright blue star does not belong to the stars of the DDO\,68.}
 Thus, the probability of a foreground star falling into the selection of
 candidates for massive stars is far from zero.
A convenient feature for identifying foreground stars is to compare the
 positions of bright stars with respect to star formation regions. The
 probability that a massive star will appear in a star cluster is always
 higher than the probability of its isolated appearance.
 A comparison of the positions of two bright blue stars in the DDO\,68 galaxy
 shows that the LBV star is indeed surrounded by a cluster of stars and gas
 clouds, while the foreground star is completely isolated. However, this
 feature is not absolutely mandatory, since bright massive stars are known
 outside the clusters.

The third factor that impedes the search is the insufficient spatial
 resolution of the telescopes. Even in our Galaxy, when using ground-based
 telescopes, it is not possible to separate close groups of massive stars
 into separate stars. Therefore, when the mass of the entire group of stars
 is attributed to a single star, errors in mass estimation can occur. The
 Hubble Space Telescope allows to separate close groups and evaluate the
 luminosities of individual stars, but it is still not enough for distant
 galaxies.

\begin{figure}
   \centering
   \includegraphics[width=11cm, angle=0]{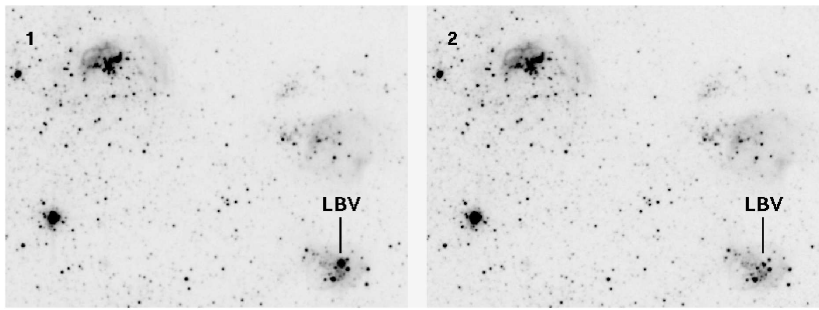}
   \caption{Images of the galactic region DDO\,68 in 2010 and 2017 in the
 F606W filter, in which the LBV star is marked by a line. A large change
 in the luminosity of the LBV star over 7 years can be seen.}
   \label{Fig3}
   \end{figure}

The unstable state of massive stars often leads to variability of their
 brightness, which can be used for search of such stars. However, the
 constancy of the brightness of a star over several years does not mean
 its stability for a longer time interval. Only long series of accurate
 photometric measurements can give an answer about brightness variability.
 Figure~\ref{Fig3} shows the images of LBV star in DDO\,68 in 2010 and 2017.
 The pictures clearly show a strong drop in its brightness from
 $V = 20\,.\!\!^{\rm m}19$ (2010) to $V = 23\,.\!\!^{\rm m}70$ (2017) (from $M_V = -10\,.\!\!^{\rm m}26$ to
$M_V = -6\,.\!\!^{\rm m}75$).

\begin{figure}
	\centering
	\includegraphics[width=7cm, angle=0]{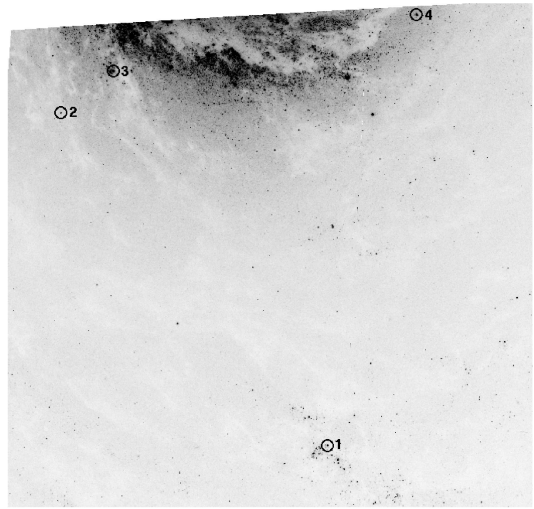}
	\caption{Part of the galaxy M\,94 in HST image in the F555W filter. The
		positions of four bright blue stars with increased brightness in the
		H${\alpha}$ filter are marked.}
	\label{Fig4}
\end{figure}
\begin{figure}
	\centering
	\includegraphics[width=7cm, angle=0]{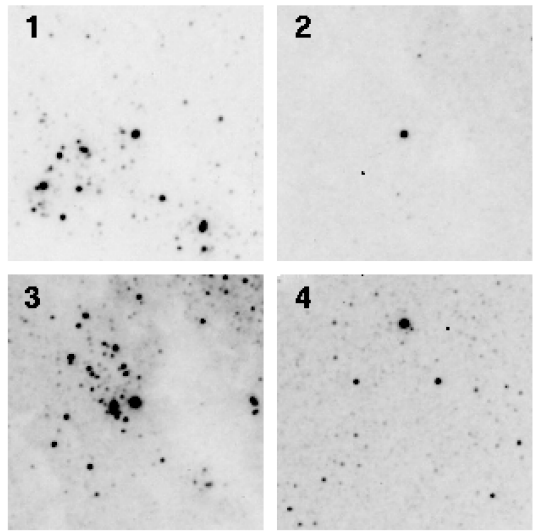}
	\caption{Regions of stars marked in Fig.~4 with numbers. Star N4 is
		shown off-center.}
	\label{Fig5}
\end{figure}
\begin{figure}
	\centering
	\includegraphics[width=5cm, angle=-90]{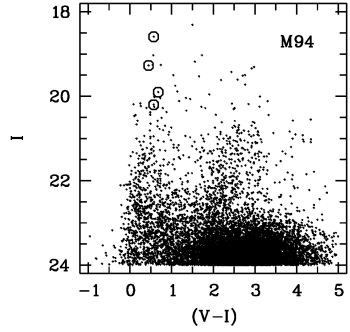}
	\caption {CM diagram of the field of the galaxy M\,94. Among the bright
		blue objects are both massive stars and possible young compact clusters.}
	\label{Fig6}
\end{figure}   
Due to the instability of massive stars, gas loss occurs and various shells
 and clouds form. In the spectra of such stars, a strong H${\alpha}$
 hydrogen emission line should be observed. The presence of the emission
 line greatly simplifies the search of candidates for massive stars. Images
 of the galaxy in H${\alpha}$ filter, together with images in other filters
 (F435W, F555W, F606W and F814W) allow to highlight bright blue stars with H${\alpha}$ emission, which
 are likely to be the sought-after massive stars. But even after such a
 selection, extraneous objects fall into the list. Young compact star
 clusters almost do not differ in profile from single stars; they have a
 strong H${\alpha}$ emission and a small color index, identical to the
 color index of bright massive stars. High resolution spectral observations
 help to separate stars from clusters, but this requires additional research.

\section{Search for massive stars in the galaxies M\,94 and NGC\,1672}
\label{sect:massive_stars}

 Figure~\ref{Fig4} shows an HST image of a part of the galaxy M\,94. The stellar
 photometry of this galaxy was performed with DOLPHOT~2.0 software 
 package\footnote{http://americano.dolphinsim.com/dolphot/dolphot.pdf}. The DOLPHOT~2.0 package
 was used in accordance with  Dolphin's recommendations 
\citep{D2016}, while
 the photometry  procedure  consisted  of  bad  pixel premasking, cosmic-ray
 particle hit removal, and further PSF photometry for the stars found in two
 filters. The photometry parameters used in our work can be found in the reference\footnote{https://cloud.mail.ru/public/Gw6v/FvP1sRmKr}. The obtained results of stellar
 photometry were selected according to the $CHI < 1.3$ and $|SHARP| < 0.3$
 parameters, which define the shape of the photometric profile of each measured
 star. This allowed us to remove all diffuse objects (star clusters, distant
 or compact galaxies) from the photometry tables, because the photometric
 profiles of these objects differed from those of stars.

After applying the DOLPHOT~2.0 package to the images of the WFC3/UVIS camera,
we get the results of VegaMag photometry, which differ from Johnson--Cousins
system. However, when DOLPHOT~2.0 photometry of the ACS camera images, we
obtain results for the stars, both VegaMag and Johnson--Cousins systems.
Therefore, it is not difficult to translate magnitudes from one system to
another. In addition, for blue LBV stars, the differences between the systems
are very small and do not even exceed the photometry error (i.e., less than
$0\,.\!\!^{\rm m}02$).
 
\begin{figure}
   \centering
   \includegraphics[width=5cm, angle=-90]{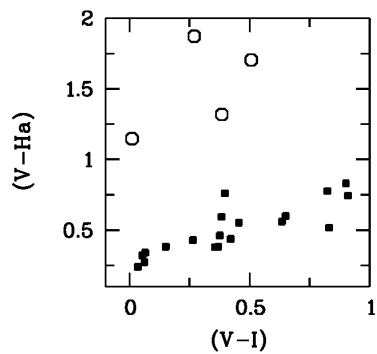}
   \caption {CM diagram of bright blue stars seen in the diagram in
 Fig.~\ref{Fig6}. Black squares denote stars without emission in the
 H${\alpha}$ line, and circles~--- stars with emission. These stars
   are marked on the image of the galaxy (Fig.~\ref{Fig4}).}
   \label{Fig7}
   \end{figure}
   \begin{figure}
   \centering
   \includegraphics[width=5cm, angle=-90]{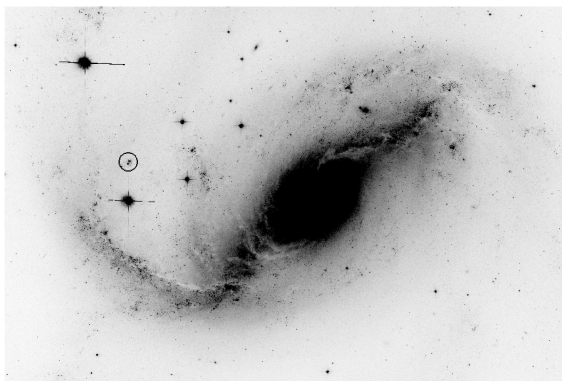}
   \caption {Galaxy NGC\,1672 in HST image in the F814W filter. The circle
 marks the position of the gas and dust nebula and star cluster containing
 stars up to $M_V = -10$.}
   \label{Fig8}
    \end{figure}
   \begin{figure}
   \centering
   \includegraphics[width=15cm, angle=0]{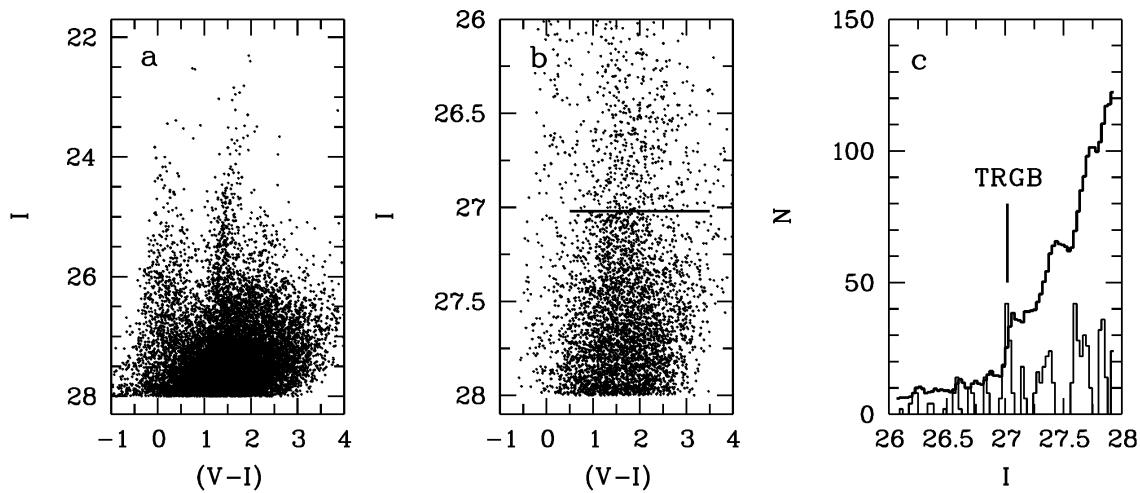}
 \caption {CM diagram of stars in the periphery of NGC\,1672 (a) and
 CM diagram of the same field (b) after the removal of stars from
 star-formation regions. The luminosity function of red giants and
 AGB stars (c) was obtained after selection stars for the color index
 $(V-I)$. The thin line shows the Sobel function, showing the increase
 in the number of stars (TRGB jump) at $I = 27\,.\!\!^{\rm m}02$.}
   \label{Fig9}
    \end{figure}
        \begin{figure}[h]
    \centering
 \begin{minipage}[t]{0.3\linewidth}
  \centering
   \includegraphics[width=45mm]{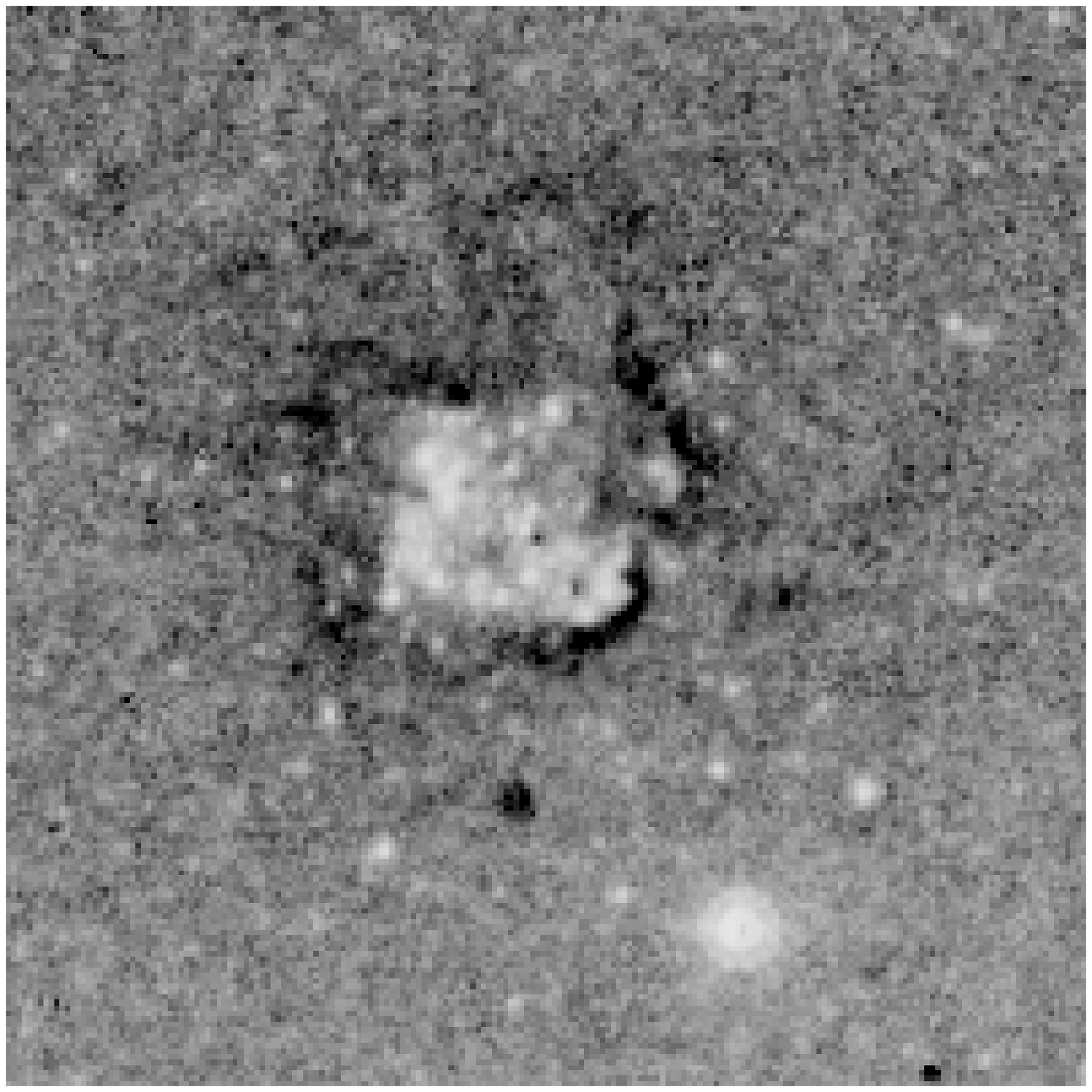}
  \end{minipage}%
  \begin{minipage}[t]{0.35\textwidth}
  \centering
   \includegraphics[width=45mm]{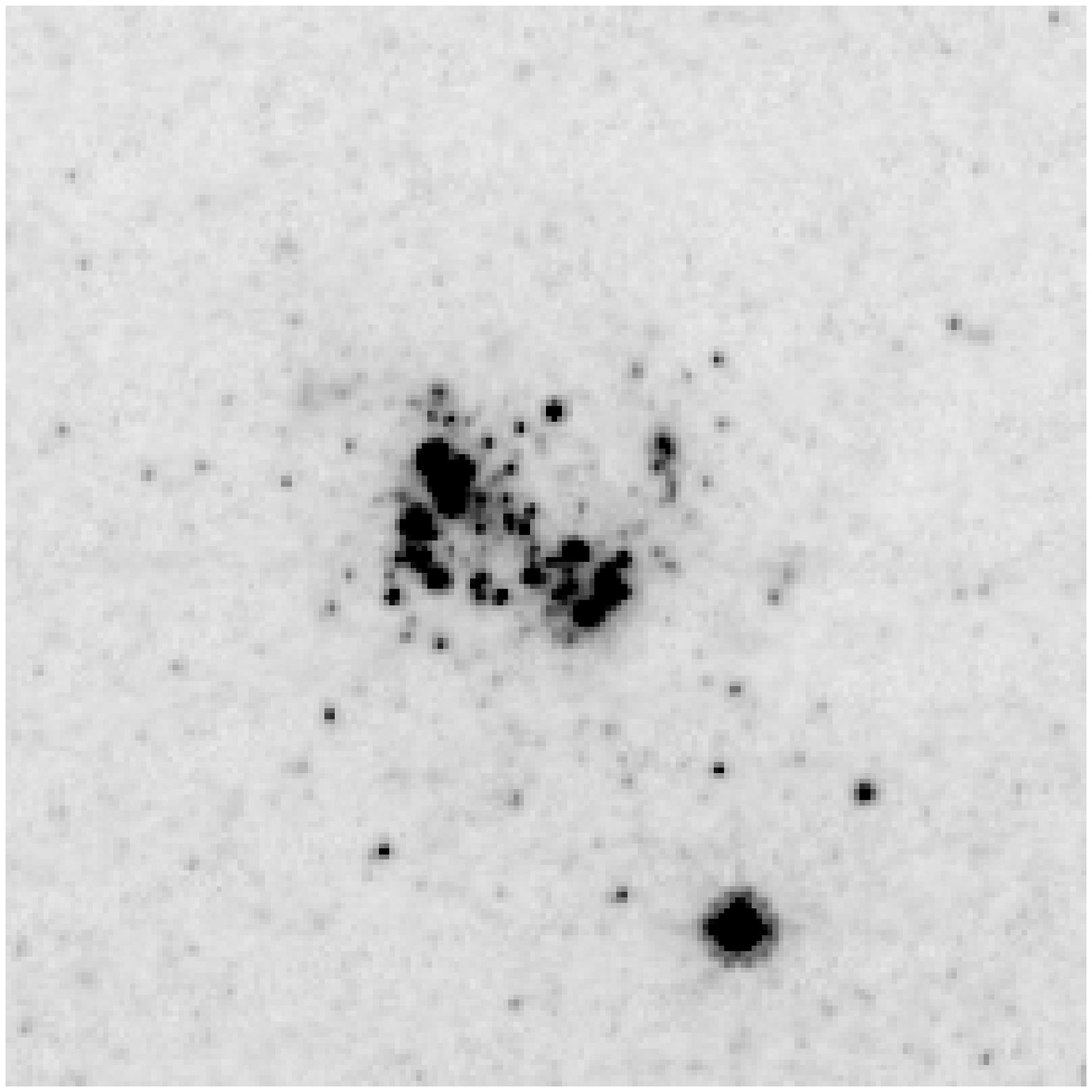}
    \end{minipage}%
  \caption{{\small On the left: nebula and star cluster cut off in
 Fig.~\ref{Fig8}. Image obtained by dividing images in the filters F658N
 and F814W. There is a visible gas shell around the cluster that has not
 dispersed due to the youth of the cluster. On the right: the same cluster
 in filter F814W. } }
  \label{Fig10}
  \end{figure}
\begin{figure}
   \centering
   \includegraphics[width=5cm, angle=-0]{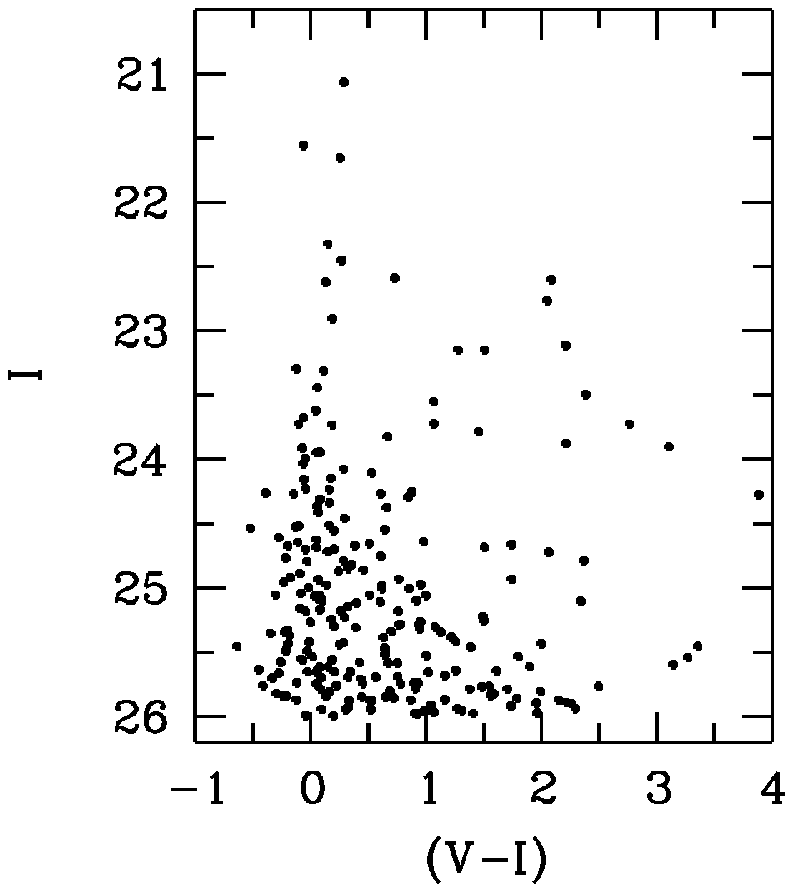}
   \caption {CM diagram of young cluster from NGC\,1672.}
   \label{Fig11}
    \end{figure}

We tested a method for identifying probable massive stars based on
photometry of galaxies in several filters. The archive HST telescope contains
images of the M\,94 galaxy (ID10402) obtained in the F435W, F555W, F658N and F814W
filters with exposures of 1200, 1000, 1400 and 900 seconds. In this galaxy,
a visual search for massive stars has already been carried out and three
probable LBV stars have been found \citep{SVF2019}.

The Hertzsprung-Russell diagram obtained after photometry of stars the M\,94
is shown in Figure~\ref{Fig6}. Bright blue stars are highlighted
on the diagram. An additional diagram using photometry in the H${\alpha}$
filter (Figure~\ref{Fig7}) makes it possible to distinguish 4 blue stars
with the possible presence of H${\alpha}$ emission lines in the spectrum.
The regions of these stsrs are shown on an enlarged scale in (Figure~\ref{Fig5}).

Stars that do not have exess in the H${\alpha}$ are marked in the diagram
(Figure~\ref{Fig7}) with black squares. They are not of interest, although
their luminocity may be even higher than that of stars with emission
in H${\alpha}$.

Spectra for three stars with H${\alpha}$ emission (Figure~\ref{Fig6}) were
obtained during the observations at the 6-m BTA \citep{SVF2019}, which
confirmed the presence of H${\alpha}$ line. These three stars change their
luminosity and can be considered as LBV stars. No spectral observations
were performed for the fourth star, but it can be assumed that this star
also belongs to the class of bright massive stars.

The result of photometric search for massive stars in M\,94 have shown that
such stars can be found using images in several filters.

 The Seyfert galaxy NGC\,1672 (Figure~\ref{Fig8}) is probably part of the
 Dorado group \citep{S1966}. Uncertainty arises from unreliable determination
 of distance. Using archival images of the Hubble Space Telescope (ID10354 and ID15654), we performed stellar photometry of this galaxy with the DAOPHOT~II \citep{Stetson_1987, Stetson_1994} and DOLPHOT~2.0 software packages.
 The selection of stars in DAOPHOT~II was carried out in the same way as in DOLPHOT~2.0: $CHI < 1.3$ and $|SHARP| < 0.3$. The resulting CM diagram of NGC~1672 stars is shown in Fig.~\ref{Fig9}a; Fig.~\ref{Fig9}b,c demonstrates the luminosity function, which marks the TRGB jump, which is necessary to determine the distance to the galaxy.

 The obtaned
 distance value ($D = 15.8 \pm0.8$~Mpc) confirms that NGC\,1672 belongs
 to the Dorado group. The obtained distance allows us to measure the energy
 of the galactic nucleus, and to determine the luminosities of the brightest
 stars, which are visible on the obtained CM diagram (Figure~\ref{Fig9}).

 Figure~\ref{Fig8}, a circle indicates a young star cluster surrounded by
 a gas shell (Figure~\ref{Fig10}) and containing stars up to $M_V = -10$.
 The CM diagram of this young cluster (Figure~\ref{Fig11}) does not differ
 from similar diagrams of other clusters. Therefore, we believe that most of
 the stars in this diagram are single stars and not cluster of stars. This
 young cluster is similar to the R136 cluster in BMO, where very massive
 stars are located.

 Whether the bright stars of the cluster NGC\,1672 are single or multiples
 can only be determined additional observations. Variability of the
 brightness of these stars would indicate that they are single stars. It
 should be noted that the stability of brightness does not prove that
 these stars are multiples or are star clusters.
 
\section{Conclusions}
\label{sect:conclusion}
The briefly described methods of searching of candidates for hypergiants
 are used by us for galaxies outside the Local Group. After compiling
 a list likely massive stars, we make spectral observation with a 6-m
 BTA telescope and obtain more detailed information about physical parameters
 of the stars. Not every galaxy contains bright hypergiants. To increase the width of the search for massive stars, we performed stellar photometry of the HST images of 320~galaxies in the northern sky. 
 Possible hypergiants have been found in 53~galaxies. Planned spectral and photometric observations with ground-based telescopes will help us to find massive stars in these galaxies.

\begin{acknowledgements}
 The study was financially supported by the Russian Foundation for Basic Research
 and the National Science Foundation of Bulgaria as a part of the scientific
 project N 19--52--18007 and Grant KP--06--Russia--9/2019.

\end{acknowledgements}

\bibliographystyle{raa}
\bibliography{tikhonov_BS}

\begin{thebibliography}{16}
\providecommand\natexlab[1]{#1}
\providecommand\JournalTitle[1]{#1}

\bibitem[{Bertelli} {et~al.}(1994)]{BBC1994}
{Bertelli}, G., {Bressan}, A., {Chiosi}, C., {Fagotto}, F., \& {Nasi}, E. 1994,
  \aaps, 106, 275

\bibitem[{Crowther} {et~al.}(2010)]{CSH2017}
{Crowther}, P.~A., {Schnurr}, O., {Hirschi}, R., {et~al.} 2010, \mnras, 408,
  731

\bibitem[{Dolphin}(2016)]{D2016}
{Dolphin}, A. 2016, {DOLPHOT: Stellar photometry}

\bibitem[{Eikenberry} {et~al.}(2004)]{EML2004}
{Eikenberry}, S.~S., {Matthews}, K., {LaVine}, J.~L., {et~al.} 2004, \apj, 616,
  506

\bibitem[{Garmany} \& {Massey}(1981)]{GM1981}
{Garmany}, C.~D., \& {Massey}, P. 1981, \pasp, 93, 500

\bibitem[{Gull}(2015)]{G2015}
{Gull}, T.~R. 2015, in Wolf-Rayet Stars, ed. W.-R. {Hamann}, A.~{Sander}, \&
  H.~{Todt}, 149

\bibitem[{Humphreys} {et~al.}(2017)]{HDH2017}
{Humphreys}, R.~M., {Davidson}, K., {Hahn}, D., {Martin}, J.~C., \& {Weis}, K.
  2017, \apj, 844, 40

\bibitem[{Pustilnik} {et~al.}(2017)]{PMP2017}
{Pustilnik}, S.~A., {Makarova}, L.~N., {Perepelitsyna}, Y.~A., {Moiseev},
  A.~V., \& {Makarov}, D.~I. 2017, \mnras, 465, 4985

\bibitem[{Shobbrook}(1966)]{S1966}
{Shobbrook}, R.~R. 1966, \mnras, 131, 365

\bibitem[{Sholukhova} {et~al.}(2018)]{SFV2018}
{Sholukhova}, O.~N., {Fabrika}, S.~N., {Valeev}, A.~F., \& {Sarkisian}, A.~N.
  2018, Astrophysical Bulletin, 73, 413

\bibitem[{Solovyeva} {et~al.}(2019)]{SVF2019}
{Solovyeva}, Y., {Vinokurov}, A., {Fabrika}, S., {et~al.} 2019, \mnras, 484,
  L24

\bibitem[{Spera} \& {Mapelli}(2017)]{SM2017}
{Spera}, M., \& {Mapelli}, M. 2017, \mnras, 470, 4739

\bibitem[{Stetson}(1987)]{Stetson_1987}
{Stetson}, P.~B. 1987, \pasp, 99, 191

\bibitem[{Stetson}(1994)]{Stetson_1994}
{Stetson}, P.~B. 1994, \pasp, 106, 250

\bibitem[{Tikhonov} {et~al.}(2014)]{TGL2014}
{Tikhonov}, N.~A., {Galazutdinova}, O.~A., \& {Lebedev}, V.~S. 2014, Astronomy
  Letters, 40, 1

\bibitem[{Yusof} {et~al.}(2013)]{YHN2013}
{Yusof}, N., {Hirschi}, R., {Meynet}, G., {et~al.} 2013, \mnras, 433, 1114

\end{thebibliography}
 
\label{lastpage}

\end{document}